\documentclass[ 
    aps,
    prl,
    10pt,
    twocolumn,
    showpacs,
    nofootinbib,
]{revtex4-2}

\usepackage{graphicx}

\usepackage{amsmath}
\usepackage{amssymb}
\usepackage{braket}

\usepackage{lipsum}
\usepackage{xcolor}
\definecolor{darkblue}{rgb}{0.0, 0.0, 0.75}
\usepackage{color}
\usepackage[colorlinks=true,
	linkcolor=darkblue,
	urlcolor=darkblue,
	citecolor=darkblue
]{hyperref}
\usepackage[english=american]{csquotes}
\usepackage[capitalize]{cleveref}

\begin{document}

    \title{Dynamical quantum phase transition with singular multipartite entanglement}
    \author{Jie Chen}
    \email {jie.chen@tu-berlin.de}
    \author{Ricardo Costa de Almeida}
    \author{Hendrik Weimer}
    \affiliation{Institut f\"ur Physik und Astronomie, Technische Universit\"at  Berlin, Hardenbergstraße 36, EW 7-1, 10623 Berlin, Germany}
    \date{\today}

    \begin{abstract}
        We investigate the nonequilibrium quench dynamics of the one-dimensional transverse-field Ising model in both integrable and nonintegrable regimes.
	    In particular, we report on a novel type of dynamical quantum phase transition (DQPT) that is characterized by a singular multipartite entanglement signature occurring at critical times in the post-quench dynamics.
        We show that this behavior is fundamentally distinct from previously studied DQPTs characterized by a nonanalytic rate function.
        We quantify the multipartite entanglement of the state by the quantum Fisher information and demonstrate that the DQPT belongs to a different universality class than the ground-state phase transition.
        Furthermore,  we perform a spectral analysis of the DQPT and demonstrate that it is a genuine nonequilibrium transition arising from the constructive interference of excited states of the system during the many-body dynamics.
        Finally, we discuss potential experimental realizations in Rydberg platforms as well as applications in the context of quantum metrology.
    \end{abstract}

    \maketitle

    \paragraph{Introduction.}
    The study of phase transitions in equilibrium and nonequilibrium systems has been one of the main driving forces in advancing our understanding of quantum many-body systems \cite{wilson1975t,fisher1998r,cardy1996s,sachdev2011q}.
    For equilibrium systems, the connection between physical properties and information-theoretic concepts has become apparent through the investigations of the entanglement structure of quantum-critical states \cite{osterloh2002s,vidal2003e,amico2008e,dechiara2018g}.
    Here, we extend this notion to dynamical quantum phase transitions occurring out of equilbirum and report a novel nonequilibrium transition that is driven by multipartite entanglement, which holds high potential for applications in quantum metrology.

    Multipartite entanglement, a generalization of the usual notion of bipartite entanglement with special relevance to quantum many-body physics \cite{horodecki2009q,dechiara2018g}, is primarily investigated in theoretical and experimental works using the quantum Fisher information (QFI) \cite{braunstein1994s,hauke2016m,pappalardi2017m,gabbrielli2018m,brenes2020m,costadealmeida2021f,menon2023m,strobel2014f,lucke2011t,lucke2014d,hales2023w,fang2025a}.
    Besides certifying the presence of multipartite entanglement, the QFI also quantifies the entanglement-induced enhancement in quantum metrology \cite{pezze2009e,hyllus2012f,toth2012m,toth2014q,pezze2018q}.
    Multipartite entanglement at quantum phase transitions manifests itself through the appearance of universal scaling laws at the critical points \cite{hauke2016m}, enabling us to use phase transitions in many-body systems as a metrological tool \cite{Gammelmark2011,Raghunandan2018,rams2018a,frerot2018q,chu2021d,gietka2022u,hotter2024c}.

    \begin{figure}
        \centering
        \includegraphics[width=\columnwidth]{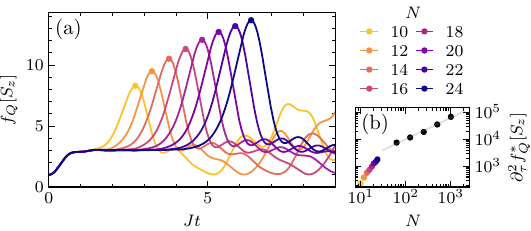}
        \caption{
	    (a) Time evolution of the QFI density \(f_Q[S_z]\) of the transverse-field Ising model following a quench to \(h/J=1\) for various system sizes.
		(b) Scaling of \(\partial_\tau^2 f^*_Q[S_z]\), the second derivative of the peak with respect to the rescaled time \(\tau\), with system size.
        }
        \label{fig:dynamics_N}
    \end{figure}

    Dynamical quantum phase transitions (DQPTs) \cite{heyl2013d,karrasch2013d,heyl2014d,canovi2014f,vajna2014d,lang2018d,cheraghi2020d,denicola2021e,corps2023t} occur when observables display nonanalytic behavior at critical times following a quench of the Hamiltonian, i.e., the dynamics is triggered via a sudden change of a coupling constant.
    In many cases, the DQPT can be understood using equilibrium concepts by relating a rate function of the many-body state to the free energy density of an equilbrium system.
    This correspondence can be made rigorous via a boundary partition function \cite{leclair1995b}, allowing to apply standard field-theoretical methods related to the renormalization group to DQPTs.
    We also note that the term \enquote{dynamical phase transition} has been employed in the literature to describe several related but distinct phenomena \cite{schuetzhold2006s,uhlmann2007v,eckstein2009t}.

    In this Letter, we introduce a new type of QFI-driven DQPT characterized by multipartite-entangled critical states emerging from out-of-equilibrium quantum many-body dynamics.
    We illustrate our results using the one-dimensional (1D) transverse-field Ising model, where the dynamics is triggered by a quench of the transverse-field strength.
    Our focus is on the production of multipartite entanglement during the time evolution.
    For quenches to the critical point, we observe that the QFI density becomes singular in time, scaling logarithmically up to characteristic system sizes, see \cref{fig:dynamics_N}.
    In contrast, for the conventional DQPT, signaled by the rate function becoming singular, we do not find a nontrivial multipartite entanglement signature.
    Vice versa, we also do not find our QFI-driven DQPTs exhibiting a singular rate function, making the two concepts fundamentally distinct.
    We further demonstrate that the scaling behavior of the QFI does not stem from the equilibrium properties.
    Instead, it originates from the constructive interference of excited states of system during the dynamics.
    Moreover, we show that the novel phase transition is not merely an artifact of the integrability of the transverse-field Ising model, as it remains robust under the inclusion of strong integrability-breaking terms to the Hamiltonian.
    Finally, we discuss potential experimental realizations of our findings, with particular emphasis on the role of dissipation and decoherence in realistic settings.

    \paragraph{Theoretical framework.}
    We consider the 1D Ising model in a transverse field for $N$ spin-\(\frac{1}{2}\) particles with periodic boundary conditions, whose Hamiltonian is given by 
    \begin{equation}
        H = 
        - J \sum_{l} \sigma_{l}^{z} \sigma_{l+1}^{z}
        - J' \sum_{l} \sigma_{l}^{z} \sigma_{l+2}^{z}
        + h \sum_{l} \sigma_{l}^{x}
        \,.
        \label{eq:model_hamiltonian}
    \end{equation}
    Here, \(\sigma_{l}^{j}\), with \(j = x,y,z\), denotes the Pauli matrices acting on the \(l\)th site, $J$ is the nearest neighbor spin-spin interaction strength, and $h$ is the strength of the transverse field.
    We also include an additional next-nearest-neighbor interaction \(J'\) in order to break the integrability of the model. 
    In the integrable limit (\(J' = 0\)) it is well known that the model has a quantum critical point at \(h/J=1.0\), which connects a ferromagnetic (\(h/J < 1.0\)) to a paramagnetic (\(h/J > 1.0\)) phase \cite{sachdev2011q}.
    To study the dynamics out of equilibrium, we prepare our system initially as a fully polarized state 
    \begin{equation}
        \ket{\psi_0} = \bigotimes_{l = 1}^{N} \frac{1}{\sqrt{2}}\left(\ket{\uparrow} -\ket{\downarrow}\right)
        \, ,
        \label{eq:initial_state}
    \end{equation} 
    which is the ground state for \(h \rightarrow +\infty\), i.e., deep in the paramagnetic phase.
    
    The QFI of the evolved state can be readily calculated, as for a pure state it is simply given by the variance
    \begin{equation}
        F_Q[S_{\vec{n}}] = 4 \left(\bra{\psi(t)}S_{\vec{n}}^2\ket{\psi(t)} - \bra{\psi(t)}S_{\vec{n}}\ket{\psi(t)}^2 \right)
        \, ,
        \label{eq:qfi_pure_state}
    \end{equation}
    where \(S_{\vec{n}}=\frac{1}{2}\sum_{l,j} n_j \sigma_{l}^{j}\) is a reference spin operator and \(\vec{n}=(n_x,n_y,n_z)\) is a vector on the Bloch sphere.
    It is worth noting that the usage of a linear operator in \cref{eq:qfi_pure_state} enables us to connect the QFI to the presence of multipartite entanglement.
    In this regard, it is convenient to use the optimal QFI direction \(\vec{o}\), defined by \(f_{Q}[S_{\vec{o}}] (t) = \max_{\vec{n}} f_{Q}[S_{\vec{n}}] (t)\), as it provides the best QFI entanglement bounds at a fixed time.

    As aforementioned, the QFI alternatively connects to the uncertainty $\Delta \phi$ of a parameter estimation protocol through the quantum Cramér-Rao bound \(\Delta \phi = 1/\sqrt{F_Q}\).
    In the specific case of a product state, we have \(F_Q[S_{\vec{o}}]\leq N\), corresponding to shot-noise scaling \(\Delta \phi \sim 1/\sqrt{N}\) in quantum metrology, whereas \(N\)-partite entangled states can reach  \(F_Q[S_{\vec{o}}] = N^2\), the Heisenberg limit scaling \(\Delta \phi \sim 1/N\) \cite{pezze2018q}.
    Within our quench protocol, the QFI \(F_Q[S_{\vec{n}}]\) corresponds to the parameter estimation of an external field \(\phi \sim S_{\vec{n}}\), which is the standard scenario in quantum metrology and arises in many instances ranging from frequency metrology to the measurement of fundamental constants such as the electric dipole moment of the electron \cite{aeppli2024c,roussy2023a}.
    
    Before proceeding, let us note that a conventional DQPT is characterized by the nonanalytic behavior of observables at critical times $t_c$.
    In previous studies \cite{heyl2013d,karrasch2013d,heyl2014d,canovi2014f,vajna2014d,lang2018d,denicola2021e,corps2023t}, this has been established via the rate function
    \begin{equation}
        \lambda = -\frac{1}{N}\log Z(t) = -\frac{1}{N}\log |\braket{\psi_0\mid\psi(t)}|^2
        \, ,
        \label{eq:rate_function}
    \end{equation}
    where the Loschmidt echo \(Z(t)\) plays the role of a partition function, while the corresponding rate function serves as an analog of the free energy density \cite{heyl2018d}.
    In order to tackle the complex correlated quantum dynamics, we employ the open-source toolbox QuTiP \cite{johansson2012q,johansson2013q,lambert2024q} to perform numerical calculations of both the QFI density $f_Q$ and the rate function $\lambda$. For systems sizes larger than $N=24$, we solve the system via a mapping onto free fermions.

    \begin{figure}
        \centering
        \includegraphics[width=\columnwidth]{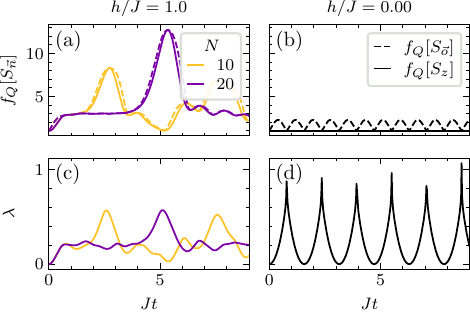}
        \caption{
			(a) Time evolution of the QFI density along the optimal (dashed) and the $z$ direction (solid) for system sizes $N=10$ and $N=20$,  \(h/J = 1\) and \(J^\prime=0.0\).
			(b) For a quench to \(h/J=0\), the QFI density remains small and smooth.
			(c,d) By contrast, the rate function $\lambda$ develops cusps in the \(h/J=0\) case, which are not present when quenching to \(h/J=1\).
			Results for the \(h/J=0\) quench are system size independent.
            See \cite{suppmat} for comparison with different scenarios.
        }
        \label{fig:dynamics_h}
    \end{figure}
 
    \paragraph{Singular QFI density.}
    \cref{fig:dynamics_h} summarizes our finding in the integrable regime, which showcases the time evolution of both the QFI density and the rate function for various system sizes for quenches to the critical point (\(h/J=1.0\)) and deep into the ferromagnetic phase (\(h/J=0.0\)).
    For both quench scenarios, the system size is fixed at \(N=20\).
    Since all spins are initially polarized along the \(x\)-axis, \(f_{Q}[S_{\vec{o}}] = 1\) at \(t = 0\) for both cases.
    It then increases for \( t>0\) as the Ising interaction introduces quantum correlations among the particles.
    For the case \(h/J= 1.0\), the QFI density initially grows slowly, followed by a sharp increase that leads to a pronounced peak at a critical time given by \(Jt_c = 5.34\).
    The corresponding QFI density increases to \(f_{Q}[S_{\vec{o}}] = 12.72\), indicating up to \(14\)-partite entanglement in the system.
    In contrast, quenching into the ferromagnet phase results in the periodic occurrence of conventional DQPTs, accompanied by oscillations of the QFI density with much smaller values (\(f_{Q}[S_{\vec{o}}] \leq 2.28\)), allowing for the certification of entanglement only up to the bipartite level.

    We further observe that for quenching (close) to the critical point, the optimal spin orientation is predominantly along the $z$-direction, and as a result, the time evolution of \(f_{Q}[S_{\vec{o}}]\) deviates only slightly from that of \(f_{Q}[S_{\vec{z}}]\) throughout the dynamics [see \cref{fig:dynamics_h} (a)].
    Noticeably, this deviation vanishes at the critical time \(Jt_c = 5.34\) and such a feature persists regardless of variations in system size. 
Based on the above observations, we define \(f^*_{Q}[S_{z}]\) as the QFI density at the peak associated with (close to) the critical value of \(h\).
Importantly, the time at which this peak occurs provides a characteristic time scale \(t_c\), that scales linearly with system size.

For very large system sizes $N\gg 32$, the peak of the QFI
  density eventually saturates. However, it maintains singular
  behavior, as can be seen in the second derivative $\partial^2
  f_Q^*[S_z]/\partial \tau^2$ with respect to the rescaled time $\tau =
  (t-t_c)/t_c$, consistent with the conventional thermodynamic
  derivative. Crucially, we show in \cref{fig:dynamics_N} (b) the second derivative diverging for system sizes up to \(N = 1024\).

    In addition to the QFI density, the dynamical behavior of the rate function \(\lambda\) also allows us to distinguish the above QFI-driven DQPT from the conventional one.
    In the case of quenching into the deep ferromagnetic phase, the rate function becomes nonanalytic at critical times \(J t_c = \pi(2n+1) /4\) with \( n= 0,1, \cdots, \) indicating that the system undergoes conventional DQPTs at those moments.
    Interestingly, in this specific case (\(h/J=0.0\)) these critical times coincide with the local minima of the QFI density in dynamics and do not scale with system sizes.
In contrast, quenching to the critical point $h/J=1$ results in a qualitatively different behavior of the rate function, with the cusps being replaced by a smooth function. This difference is particularly stark when integrability is broken, which we will discuss below. Specifically, we refer to the supplemental material \cite{suppmat} for additional data constrasting the dynamics in the QFI-driven DQPT and conventional DQPT cases.
    
    Those distinctions suggest that the QFI-driven DQPT observed above is of a different nature from the previously studied DQPTs characterized by nonanalytic rate function.
    However, we believe that it is an appropriate generalization of the concept of DQPT, since the singular QFI plays the role of a diverging susceptibility [see Eq.~(\ref{eq:qfi_pure_state})] and it exhibits universal scaling behavior (see below) - two hallmark features of phase transitions both in and out of equilibrium.
    To substantiate our claim, we further examine the scaling behavior of \(f^*_{Q}[S_{z}]\) in \cref{fig:dynamics_N}.
    Our analysis confirms that \(f^{*}_{Q}[S_{z}]\) scales logarithmically up to a characteristic size of the system, and becomes singular in time.
    These observations form our main result: a new type of nonequilibrium phase transition signaled by a singular QFI density.

    \paragraph{Genuine nonequilibrium transition.}
    In the following, we demonstrate that the nonanalytic QFI density is driven by the constructive interference of excited states of the system during the many-body dynamics, a genuinely nonequilibrium effect that cannot be explained by the static properties of the post-quench Hamiltonian.
    To this end, we employ the eigenbasis of the Hamiltonian \(H=\sum_n E_n \ket{\phi_n}\bra{\phi_n}\) and express the time-evolved state as \(\ket{\psi (t)} = \sum_n C_{n}  e^{-i E_{n} t} \ket{\phi_{n}}\), where \(C_n = \braket{\phi_n\mid \psi_0}\).
    In this way, the QFI density can always be decomposed as \(f_{Q}[S_{z}] = f_{Q}^{D}[S_{z}] + f_{Q}^{O}[S_{z}]\).
    Here, \(f_{Q}^{D}[S_{z}]\) denotes the diagonal, time-independent contribution, while \(f_{Q}^{O}[S_{z}]\) represents the off-diagonal, time-dependent part.
    Given the initial state $\ket{\psi_0}$ in Eq.~\eqref{eq:initial_state}, the diagonal and off-diagonal contributions to the QFI density can be written as 
    \begin{align}
        f_{Q}^{D}[S_{z}] &= \frac{1}{N}\sum_{n} C_{n}^{*} C_n [S_{z}^{2} ]_{nn} \, , \nonumber \\
        f_{Q}^{O}[S_{z}] &= \frac{1}{N}\sum_{m \neq n} C_{m}^{*} C_{n} e^{-i \omega_{m n} t} [S_{z}^{2} ]_{m n} \, ,
        \label{eq:qfi_diag_offdiagonal}
    \end{align}
    where \([S_{z}^{2} ]_{m n} = \braket{\phi_m\mid  S_z^{2} \mid \phi_n}\) and \( \omega_{mn} = E_n - E_m\).
    \Cref{eq:qfi_diag_offdiagonal} holds due to the parity symmetry of the Hamiltonian \(H \), which is preserved during time evolution.
    Since the initial state \(\ket{\psi_0}\) lies in the even-parity sector, matrix elements such as \([S_{z} ]_{m n}\) vanish.

    \begin{figure}[b]
        \centering
        \includegraphics[width=\columnwidth]{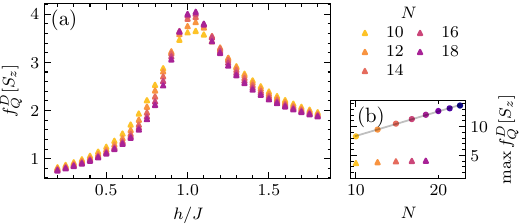}
        \caption{
	    (a) Diagonal contributions to the QFI density \(f^*_Q[S_z]\) as a function of the post-quench value \(h\) for different system sizes.
	    (b) Scaling of the QFI peak as a function of system size, showing a qualitative difference between the scaling total QFI (circles) and the diagonal contributions (triangles), which does not scale.}
        \label{fig:diagonal}
    \end{figure}
    Physically, the diagonal contribution is equivalent to the infinite time-averaged value of the QFI density and does not scale with system size in the thermodynamic limit \cite{pappalardi2017m}.
    We explicitly show in \cref{fig:diagonal} that the value of \(f_{Q}^{D}[S_{z}]\) remains bounded as the system size increases, and therefore cannot account for the scaling of the QFI density.
    Thus, our novel DQPT is a genuine nonequlibrium effect, originating from off-diagonal correlations among excited states.
    We also note that our findings are consistent with the framework of the eigenstate thermalization hypothesis (ETH).
    In that setting, the observables are averaged over an infinite time \cite{gogolin2016e,abanin2019c}, while our DQPT manifests itself only at isolated critical times $t_c$.
    Hence, the singularity vanishes under time averaging, and the ETH behavior is recovered.

    It is also worth noting that, albeit sharing the same critical value (\(h/J=1\)) with the ground state phase transition \cite{hauke2016m}, our novel DQPT cannot be attributed to the QFI divergence of the ground state.
    This conclusion follows from the previous discussion of the diagonal contribution, which bounds the ground state contribution.
    In particular, the decay of the ground state weight \(|C_0|^2\) is sufficient to suppress the ground state QFI density divergence.

    \begin{figure}
        \centering
        \includegraphics[width=\columnwidth]{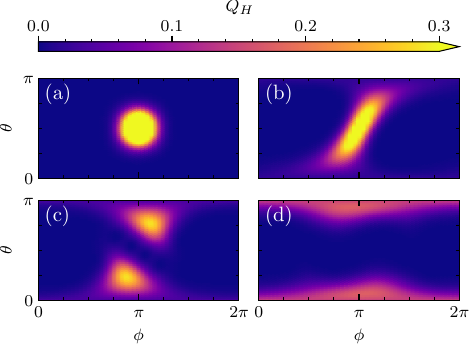}
        \caption{
	    Snapshots of the Husimi distributions at (a) $Jt = 0$,  (b) $Jt = 0.5$, (c) $Jt = 4.4$ and (d) $Jt = 5.34$ following a quench to \(h/J = 1\).
	    The final time corresponds to the critical time, where the distribution becomes strongly concentrated close to the poles of the Bloch sphere, corresponding to a GHZ-like state.
            Data is shown for $N = 20$ with a color scale cutoff for improved visibility of the main features.
        }
        \label{fig:husimi}
    \end{figure}

    \paragraph{Dynamical formation of macroscopic GHZ-like states.}
    Apart from the QFI singular behavior, another interesting feature of our novel DQPT is the dynamical formation of states closely resembling a Greenberger-Horne-Zeilinger (GHZ) state $(\ket{\uparrow}^{\otimes N} + \ket{\downarrow}^{\otimes N})/\sqrt{2}$.
    To illustrate this, we employ the Husimi distribution \(Q_{H} (\theta, \phi,t) = \frac{N+1}{4\pi}  |\braket{\psi(t) \mid \theta, \phi }|^2\) where \(\ket{\theta, \phi}\) are spin coherent states, that is, product states of spins polarized along the same direction \(\vec{n} = (\sin\theta \cos\phi, \sin\theta \sin \phi, \cos \phi)\) \cite{arecchi1972a,radcliffe1971s,zibold2010c,tomkovic2017e}.
    
    \Cref{fig:husimi} shows the time evolution of \(Q_{H} (\theta, \phi,t)\) for \(N = 20\) and \(h/J = 1.0\).
    Since the system is initially prepared in a product state, the distribution exhibits a single pronounced peak in \cref{fig:husimi} (a).
    As time evolves, the spin-spin interactions inject quantum correlations into the system and lead to the squeezing of the Husimi distribution [see Fig.~\ref{fig:husimi} (b)].
    This behavior corresponds to the rapid growth of the QFI value within \(Jt \in [0,0.5]\), as seen in \cref{fig:dynamics_N} (a).
    Afterwards, the distribution does not change significantly (results not shown), corresponding to the QFI plateau for \(Jt \in[0.5, 4.4]\) in \cref{fig:dynamics_N} (a).
    At \(Jt \approx 4.4\), two separate regions form and they move towards the north and south pole as time evolves.
    Finally, at the critical time \(Jt_c = 5.34\), these two regions separate completely and the corresponding QFI value reaches its maximum \(f^{*}_{Q}[S_{z}]\).
    Importantly, at the critical time, the Husimi distribution is highly concentrated around the two poles of the Bloch sphere, closely resembling that of a GHZ state, where it reduces to two singular points at the poles.
    We also note that the metrological usefulness of the critical state is very similar to the GHZ state \cite{suppmat}.

    \begin{figure}
        \centering
        \includegraphics[width=0.9\columnwidth]{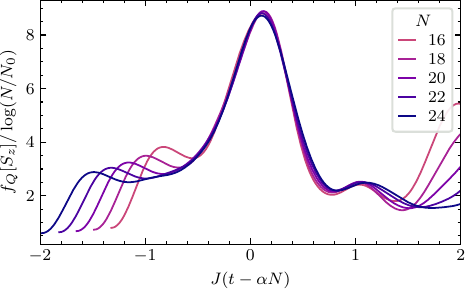}
        \caption{
	    Rescaled dynamics of the QFI density for the nonintegrable case with \(J'=J\), showing a data collapse close to the critical time \(t_c\).
	    Outside the critical regime, given by \(J|t-t_c|\lesssim 1/2\), the data collapse breaks down.
	    The nonuniversal constants \(N_0\), \(\alpha\) and \(\beta\) describe the critical QFI density, \(f^*_Q[S_z] \propto\log (N/N_0)\), and the critical time, \(t_c =\alpha N+\beta\).
	    They depend on the microscopic details of the system, such as the ratio \(J'/J\).
	}
	\label{fig:collapse}
    \end{figure}

    \paragraph{Breaking integrability and universality.}
    One special feature of the transverse-field Ising model is its integrability, as it can be mapped onto a system of free Bogoliubov fermions \cite{sachdev2011q}.
    Hence, it is natural to ask whether our novel DQPT is merely an artifact of integrability, or whether it persists when integrability is broken.
    To this end, we utilize the same protocol as in the integrable limit and monitor the time evolution of the QFI when integrability is broken by introducing a next-nearest-neighbor interaction \(J'\).
    We also use the nonintegrable model to test for universal scaling behavior, one of the hallmarks of a phase transition.
    Specifically, it is expected that close to the critical point, all deviations should collapse into a single function \cite{cardy1996s}, either due to a shift from the critical time or a finite system size.
    To verify this we evaluate the QFI density at the critical point, which is shifted for \(J'\neq 0\), as the effective strength of the ferromagnetic interaction is increased.
    \Cref{fig:collapse} depicts our results for a quench with \(J'=J\) and \(h/J \approx 2.475(5)\), the corresponding critical point, and shows a data collapse for the rescaled data.
    This unambiguously demonstrates that our novel DQPT is not merely a consequence of the integrability of the model and persists deep into the nonintegrable regime, consistent with the principle of universality.

    \begin{figure}
        \centering
        \includegraphics[width=\columnwidth]{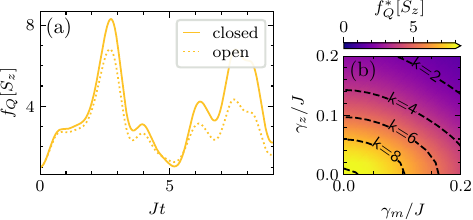}
        \caption{
            QFI density in an open system for \(N=10\) and \(J^\prime = J\).
	    (a) Time evolution of \(f_{Q}[S_{z}]\) for \(N = 10\) in a closed system (solid line) and an open system with $\gamma_z/J =\gamma_m/J = 0.05$ (dotted line).
	    (b) Boundaries of $k$-partite entanglement.
	    Over a large range of decoherence and dissipation parameters, up to 8-partite entanglement is found.
    	}
    	\label{fig:open}
    \end{figure}

    \paragraph{Experimental realization and incoherent dynamics.}
    Let us now turn to possible experimental realizations of our novel DQPT.
    Most prominently, the Ising model in a transverse field has recently been implemented in experiments with ultracold Rydberg atoms and trapped ions \cite{Bernien2017,Zhang2017,jurcevic2017d}.
    Both realizations offer to tune the strength of the next-nearest-neighbor term $J'$ over a wide range.
    In addition to their tunability, any realistic implementation will also have to deal with residual interactions with the environment.
    Hence, it is worthwhile to study the influence of dissipation and decoherence on the QFI density.
    Specifically, we assume the system is governed by the quantum master equation \cite{breuer2007t}
    \begin{align}
        \frac{d \rho}{dt}  &= -i [H, \rho] + \gamma_z \sum_l (\sigma_{l}^{z} \rho \sigma_{l}^{z} - \rho) \\ \nonumber
        & + \gamma_m \sum_l (\sigma_{l}^{-} \rho \sigma_{l}^{+} - \sigma_{l}^{+} \sigma_{l}^{-} \rho/2 - \rho \sigma_{l}^{+} \sigma_{l}^{-}/2)
        \, .
    \end{align}
    Here $\rho$ is the system's density matrix and $\gamma_z$ and $\gamma_m$ control the strength of decoherence and dissipation, respectively.
    In the case of ultracold Rydberg systems, the transverse field is typically on the order of $h/\hbar= 2\pi \times 1\,\text{MHz}$, while decoherence and dissipation are about two to three orders of magnitudes weaker \cite{Bernien2017}.
    As a conservative estimate, we choose $\gamma_z/J = 0.05$ and $\gamma_m/J = 0.05$ and depict the QFI dynamics in \cref{fig:open} (a), using a more general expression for the QFI for mixed states \cite{braunstein1994s}.
    Due to the impact of the external environment, the first peak of the QFI density is only slightly reduced, while later peaks show a more pronounced reduction.
    However, even for our conservative estimate on the strength of decoherence and dissipation, the central feature of the novel DQPT remains clearly visible and large multipartite entanglement persists over a wide range of parameters [see \cref{fig:open} (b)].
    Finally, we note that the maximum of the QFI density can be experimentally detected by measuring the variance of the $S_z$ operator \cite{suppmat}.

    \paragraph{Conclusion.}
    In summary, we have presented a novel dynamical phase transition characterized by singular multiparticle entanglement dependence on time.
    Despite the scaling of the quantum Fisher information being logarithmic and limited by a characterisic system size, we observe large values of multipartite entanglement, which are also remarkably robust against decoherence and dissipation.
    This combination makes our novel DQPT a very promising candiate to realize a quantum enhancement in metrological applications.
    While we laid our focus on the Ising model in a transverse field, our findings can be expected to be readily generalized to other spin models \cite{zunkovic2016d,perez-garcia2024d} or to cavity QED systems \cite{su2024q,saleem2025q}.

    \begin{acknowledgments}
	We thank L.~Mathey, L.~You and X. H.~Liang for fruitful discussions.
	Financial support by Berlin Quantum and by the German Federal Ministry of Research, Technology and Space through the project \enquote{Quantencomputer mit gespeicherten Ionen für Anwendungen (ATIQ)} (Subproject No. 13N16612) is gratefully acknowledged.
    \end{acknowledgments}

    \bibliographystyle{myaps}
    \bibliography{references}

\end{document}